\documentclass[a4paper]{jpconf}
\usepackage{graphicx}
\usepackage{pgf}
\usepackage[greek,british]{babel}
\newcommand{\gr}[1]{{\foreignlanguage{greek}{#1}}} 
\newcommand{\particle}[1]{\textnormal{#1}} 
\newcommand{\partgr}[1]{{\particle{\gr{#1}}}} 
\newcommand{\pion}{\partgr{p}}
\newcommand{\lamb}{\partgr{L}}

\begin{document}
\title{Correlations and flavors in jets in ALICE}

\author{Filip Krizek for the ALICE collaboration}

\address{Nuclear Physics Institute, Hlavn\'\i{} 130, \v{R}e\v{z}, Czech Republic}

\ead{filip.krizek@cern.ch}

\begin{abstract}
We report on the measurement of hadron composition in charged jets in pp at $\sqrt{s}=7$~TeV
and show the first data on particle type dependent jet fragmentation at the LHC.
Further, we present $(\lamb+\bar{\lamb})/2\particle{K}^{0}_{\mathrm{S}}$ 
ratios measured  in charged jets 
in Pb--Pb collisions at $\sqrt{s_{\mathrm{NN}}} = 2.76$~TeV 
and  in \particle{p}--Pb collisions at $\sqrt{s_{\mathrm{NN}}} = 5.02$~TeV.
While the ratio of the inclusive $p_{\mathrm{T}}$ spectra of  $\lamb$ and  
  \particle{K}$^{0}_{\mathrm{S}}$
  exhibits centrality dependent enhancement both in Pb--Pb and \particle{p}--Pb system, the
$(\lamb+\bar{\lamb})/2\particle{K}^{0}_{\mathrm{S}}$ 
  ratio measured in charged jets reveals that jet fragmentation does not contribute to the observed baryon anomaly.

Finally, we discuss  the measurement of semi-inclusive $p_{\mathrm{T}}$ spectra of charged jets 
that recoil from a high-$p_{\mathrm{T}}$ hadron trigger in Pb--Pb
 and \particle{p}\particle{p} collisions at $\sqrt{s_{\mathrm{NN}}} = 2.76$~TeV and $\sqrt{s} = 7$~TeV,  respectively.
 The jet yield uncorrelated with the trigger hadron is removed at the event-ensemble level
 without introducing a bias on the jet population which is therefore infrared and collinear safe.
 The recoil jet yield in central Pb--Pb is found to be suppressed 
  w.r.t. that from \particle{p}\particle{p} PYTHIA reference.
 On the other hand, there is no sign of intra-jet broadening even for anti-$k_{\mathrm{T}}$ jets 
  with a resolution parameter as large as $R=0.5$.

\end{abstract}

\section{Introduction}
A jet is a collimated spray of particles that is produced in the 
 process of parton showering 
 when a  highly virtual parton fragments. 
Since it is ambiguous to associate final state hadrons with initial partons,
  the assignment of  hadrons to jets is based on 
the decision of reconstruction algorithm.
Jet reconstruction algorithms are designed to recover the four-momentum of the initial parton 
by summing up momenta
 of  particles in the final state. In order to have a clear correspondence  between jets in theory and experiment,
 jet algorithms have to be infrared and collinear safe \cite{ex1}.

Jets have several useful properties which predestine them as a 
convenient probe to study the medium  created in a collision of two relativistic nuclei.
First of all, we have a good understanding  of jet production on perturbative
 QCD level in elementary reactions.
Further, it is known that processes with large four-momentum transfer $Q^2$
 occur in the initial stage of the nucleus-nucleus collision before the 
quark gluon plasma is formed. Hard-scattered partons then interact with 
the medium of deconfined quarks and gluons and loose energy.
This is manifested by the so-called jet quenching phenomenon when a jet pair with a high transverse momentum
imbalance is observed \cite{at1,cms1}.
The principal goal why the jet-medium interaction is studied is to understand 
the nature of  medium-induced parton energy loss mechanism
 and its possible connection to the strongly coupled limit of QCD.


\section{Jets in ALICE}
A detailed description of the ALICE detector can be found in \cite{ex2}.
Here let us just briefly mention that precise tracking is 
provided by a six-layered silicon vertex tracker in  combination with
 a large Time Projection Chamber (TPC).
Both detectors are placed in the core of ALICE central barrel where the solenoidal magnetic field reaches 0.5~T.
ALICE  has nearly uniform acceptance and efficiency coverage for tracks in the pseudorapidity range $|\eta|<0.9$ 
 in the full azimuth throughout a wide transverse momentum range 
from $\approx 150$~MeV/$c$ to $100$~GeV/$c$.
ALICE has also partial coverage by the electromagnetic calorimeter EMCal which covers  $107$~deg
 in azimuth and $|\eta|<0.7$ in pseudorapidity. 

Jets are reconstructed either from tracks only (\emph{charged jets})
or by combining tracks and calorimeter clusters (\emph{full jets}).
Whereas charged jets can be reconstructed in the full azimuth,
acceptance of full jets is restricted by the incomplete coverage of the electromagnetic calorimeter.
In order to remove partially reconstructed jets that appear at the boarder of 
the ALICE acceptance,
the jet pseudorapidity range is limited  by means of a fiducial cut, 
namely for charged jets we use $|\eta_{\mathrm{jet}}|<0.9-R$,
where $R$ is the resolution parameter of the given jet algorithm. 
Jets are reconstructed using the infrared 
and collinear safe anti-$k_{\mathrm{T}}$ and $k_{\mathrm{T}}$ algorithms \cite{ca0}.

Soft bulk hadrons created in a heavy-ion collision 
constitute a significant background for jet reconstruction
  and cause a shift in the jet energy scale. 
The transverse momentum of reconstructed jets $p_{\mathrm{T,jet}}^{\mathrm{raw}}$
 is therefore  corrected for the average contribution of the bulk hadrons 
using the formula
$$ p_{\mathrm{T,jet}} = p_{\mathrm{T,jet}}^{\mathrm{raw}} - \rho A_{\mathrm{jet}}$$
where $A_{\mathrm{jet}}$ denotes jet area and $\rho$ is the mean background density. 
The mean background density is estimated on an event-by-event basis employing
the standard area based method \cite{ca1}.
The finite width of the jet energy resolution is given by two sources. 
The first source of jet energy smearing are the local background fluctuations that occur in each event.
The second  contributor to the jet energy resolution are detector effects.
The response matrix that relates the $p_{\mathrm{T}}$ of reconstructed and true jets 
  is assumed to factorize into a product of matrices that  
describe momentum smearing by local background fluctuations and momentum smearing by detector effects.
The matrix is inverted by means of common 
regularized unfolding techniques based on the SVD decomposition \cite{HO} or Bayes' theorem \cite{AG}
implemented in the RooUnfold package \cite{Ad}.

\section{ Hadron structure of charged jets in pp collisions at $\sqrt{s}=7$~TeV}
Data on hadron composition of jets provides an important benchmark for 
the theory and is also needed for
 the fine tuning of the commonly used event generators such as PYTHIA \cite{PY}.
ALICE has measured the inclusive $p_{\mathrm{T}}$ spectra of hadrons
in charged jets in pp collisions at $\sqrt{s}=7$~TeV \cite{Ch1}.
To extract the yields of identified particles  
in a given track $p_{\mathrm{T}}$ bin, we use
the ionization energy loss  $\mathrm{d}E/\mathrm{d}x$ measured by the ALICE TPC.
The fractions of the most abundant particle species, 
which are $\pion$, \particle{K}, \particle{p} and \particle{e},
are extracted using two methods. The first one parameterizes the $\mathrm{d}E/\mathrm{d}x$
distribution corresponding to individual particle species 
 in a given  track $p_{\mathrm{T}}$ bin
with Gaussian functions that have  
the width and mean constrained by an analytic model \cite{Ch2,Ch3}.
The second approach substitutes the Gaussian 
functions with data driven templates of particle 
$\mathrm{d}E/\mathrm{d}x$ \cite{Ch4}.  Both models provide consistent results.

\begin{figure}[ht]
\includegraphics[width=36pc]{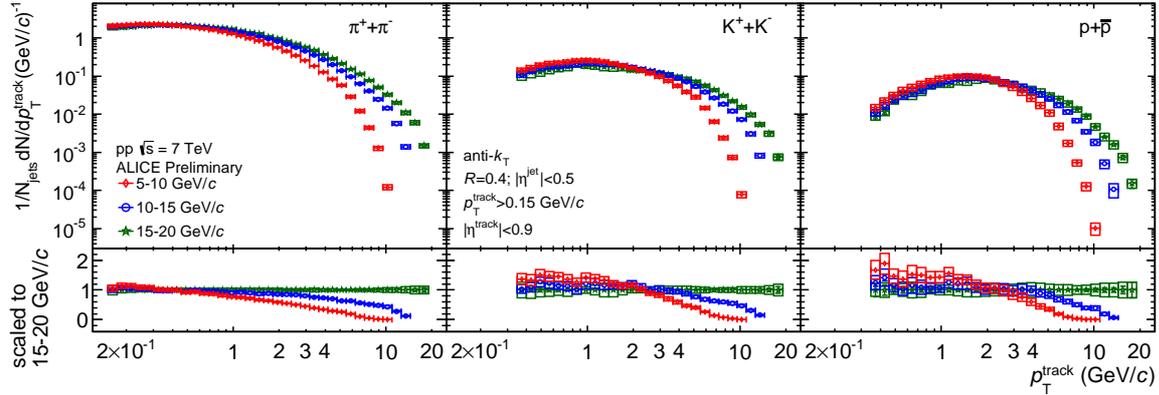}\hspace{2pc}%
\begin{minipage}[b]{36pc}\caption{\label{Y1} 
 $p_{\mathrm{T}}$ spectra of $\pion$, \particle{K}, and \particle{p}
in anti-$k_\mathrm{T}$ jets with $R=0.4$ in \particle{pp} collisions at $\sqrt{s}=7$~TeV.
The spectra are shown for three jet transverse momentum bins: 5--10~GeV/$c$, 10--15~GeV/$c$,
and 15--20~GeV/$c$. 
}  
\end{minipage}
\end{figure}

In Fig.~\ref{Y1} we present the fully corrected $p_{\mathrm{T}}$ spectra 
of pions, kaons and protons in charged anti-$k_\mathrm{T}$ jets with $R=0.4$.
This data is the first measurement of particle type dependent jet fragmentation at the LHC.
 Let us just briefly mention that 
the relative abundance of \particle{K} w.r.t. \pion\  grows 
with the increasing fraction of $p_{\mathrm{T}}$  carried  by the particle in 
the jet. 
On the other hand, with the increasing fraction of $p_{\mathrm{T}}$  carried
 by the jet constituent, protons 
are getting suppressed w.r.t. pions which signalizes the suppression of the leading baryon
 in the jet.

\section{ $\lamb/\particle{K}^{0}_{\mathrm{S}}$ ratio in charged jets in Pb--Pb and p--Pb }
In \cite{ex3}, PHENIX reported that
the ratio of the inclusive $p_{\mathrm{T}}$ spectra of $\particle{p}$ to $\pion$
exhibits a strong centrality dependence
 in Au--Au collisions at $\sqrt{s_{\mathrm{NN}}}=200$~GeV. A similar behavior of the ratio was
observed also for other baryon to meson ratios, e.g., $\lamb/\particle{K}^{0}_{\mathrm{S}}$ 
\cite{ex4}.  Hence, this effect is sometimes referred to as the baryon anomaly. 

\begin{figure}[ht]
\begin{minipage}{18pc}
\includegraphics[width=18pc]{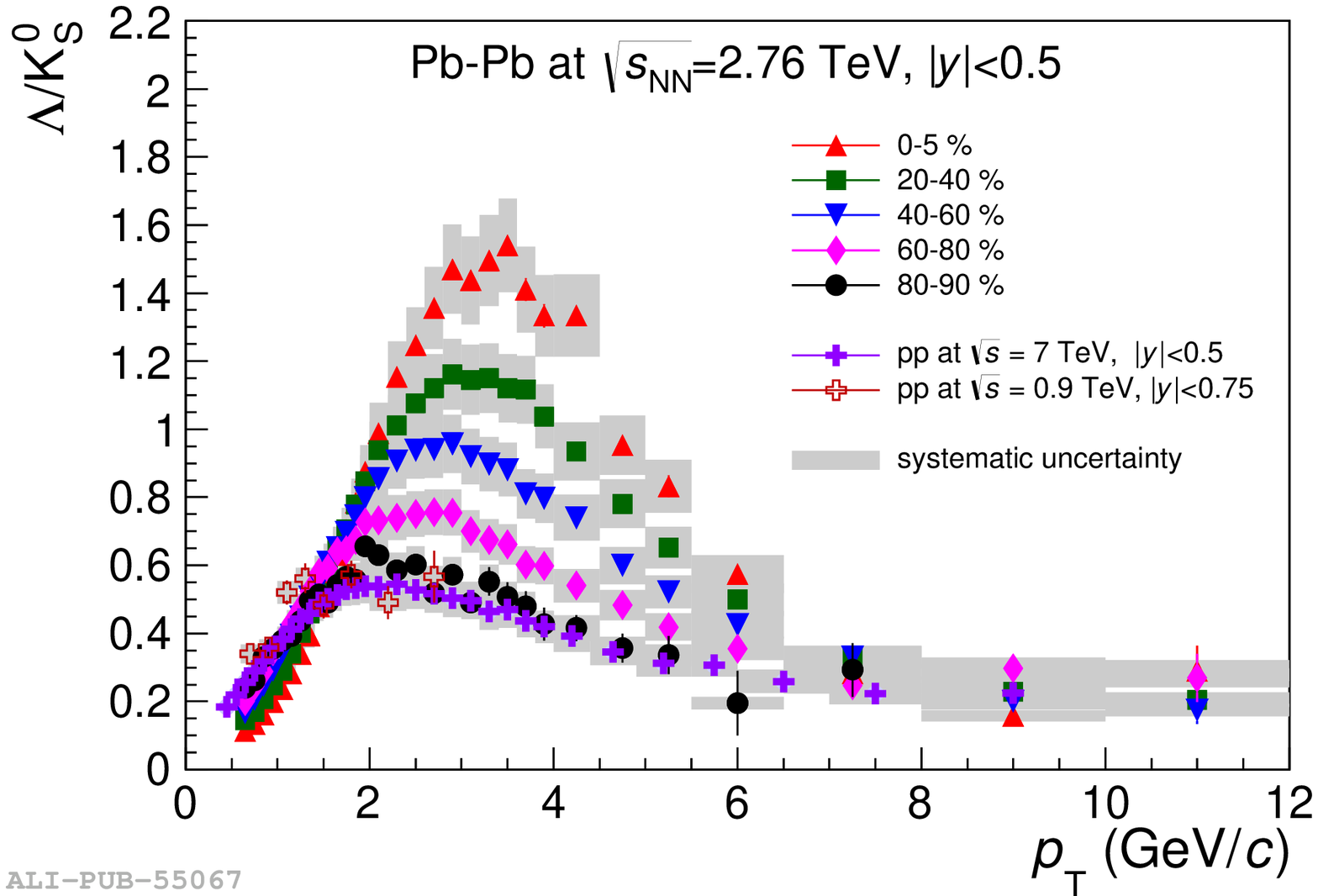}
\caption{\label{FigLKs1} Ratio of inclusive $p_{\mathrm{T}}$ spectra of $\lamb$ and $\particle{K}^{0}_{\mathrm{S}}$ measured in Pb--Pb and \particle{pp} collisions.  Centrality bins used in Pb--Pb
are quoted in legend.  For more details see \cite{A1}. }
\end{minipage}\hspace{2pc}%
\begin{minipage}{18pc}
\includegraphics[width=18pc]{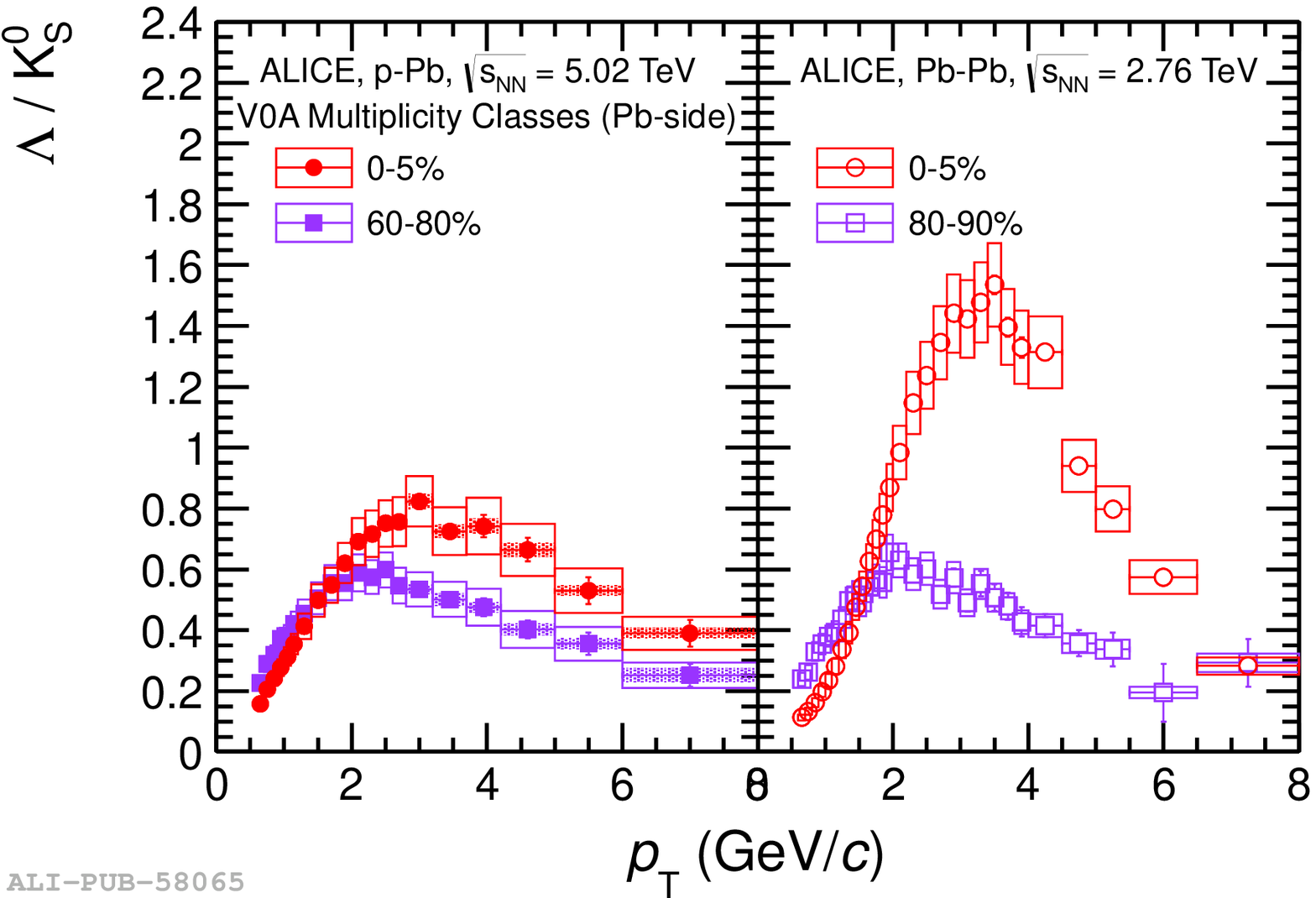}
\caption{\label{FigLKs2} Ratio of inclusive $p_{\mathrm{T}}$ spectra of $\lamb$ and 
$\particle{K}^{0}_{\mathrm{S}}$ measured in p--Pb (left) 
and Pb--Pb (right) collisions.
The corresponding centrality selections 
are quoted in the legend. See \cite{A2} for more details.  }
\end{minipage} 
\end{figure}

\begin{figure}[ht]
\begin{minipage}{18pc}
\includegraphics[width=18pc]{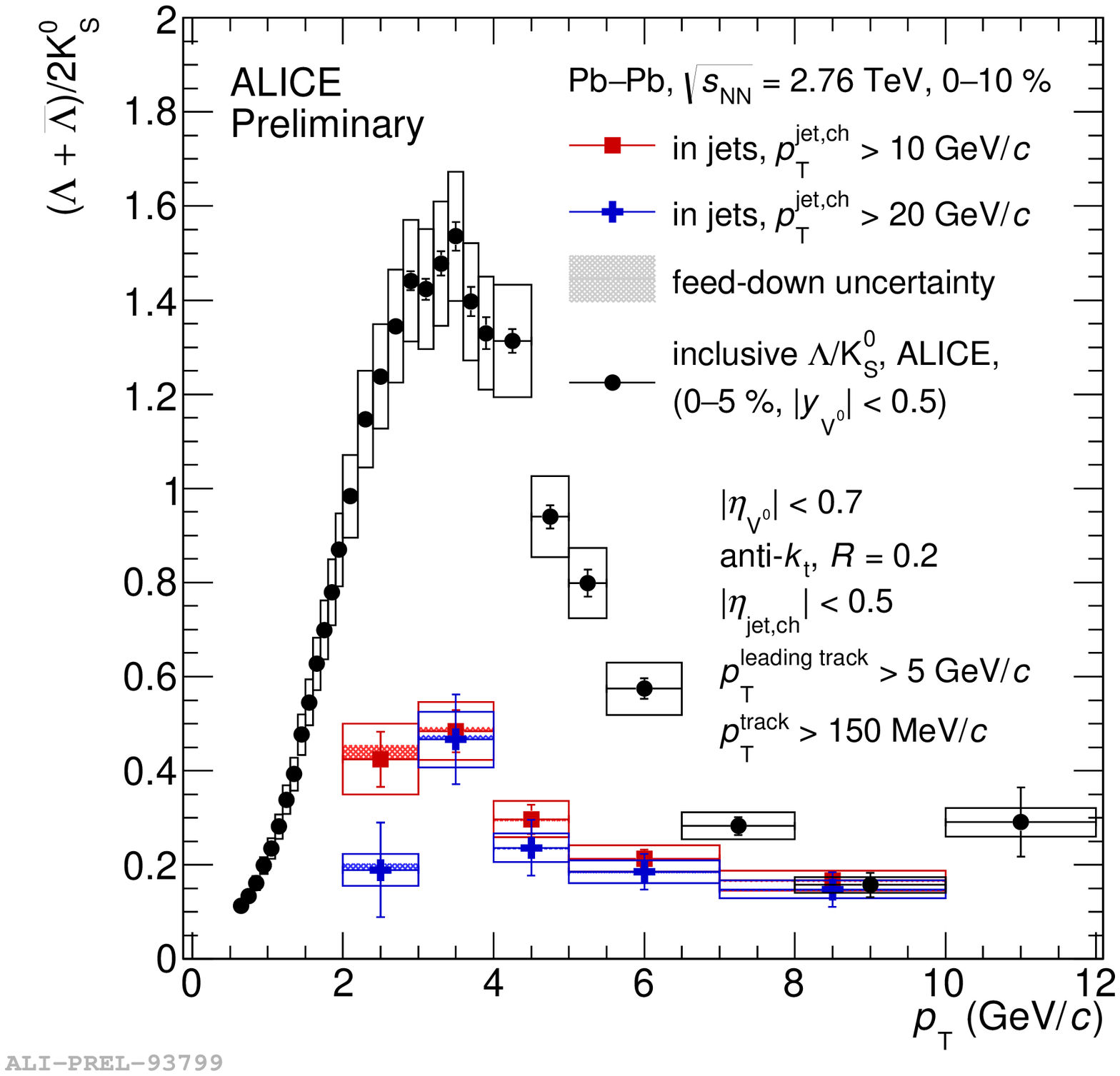}
\caption{\label{FigLKs3}  Ratio of $p_{\mathrm{T}}$ spectra of $\lamb+\bar{\lamb}$ and $\particle{K}^{0}_{\mathrm{S}}$ measured in 0--10\% most central Pb--Pb collisions at $\sqrt{s_{\mathrm{NN}}}=2.76$~TeV.
Black circles show the ratio obtained from  inclusive spectra, red squares and blue crosses
correspond to the ratios measured in charged jets with transverse momentum 
larger than 10 GeV/$c$ and 20 GeV/$c$, respectively.
}
\end{minipage}\hspace{2pc}%
\begin{minipage}{18pc}
\includegraphics[width=18pc]{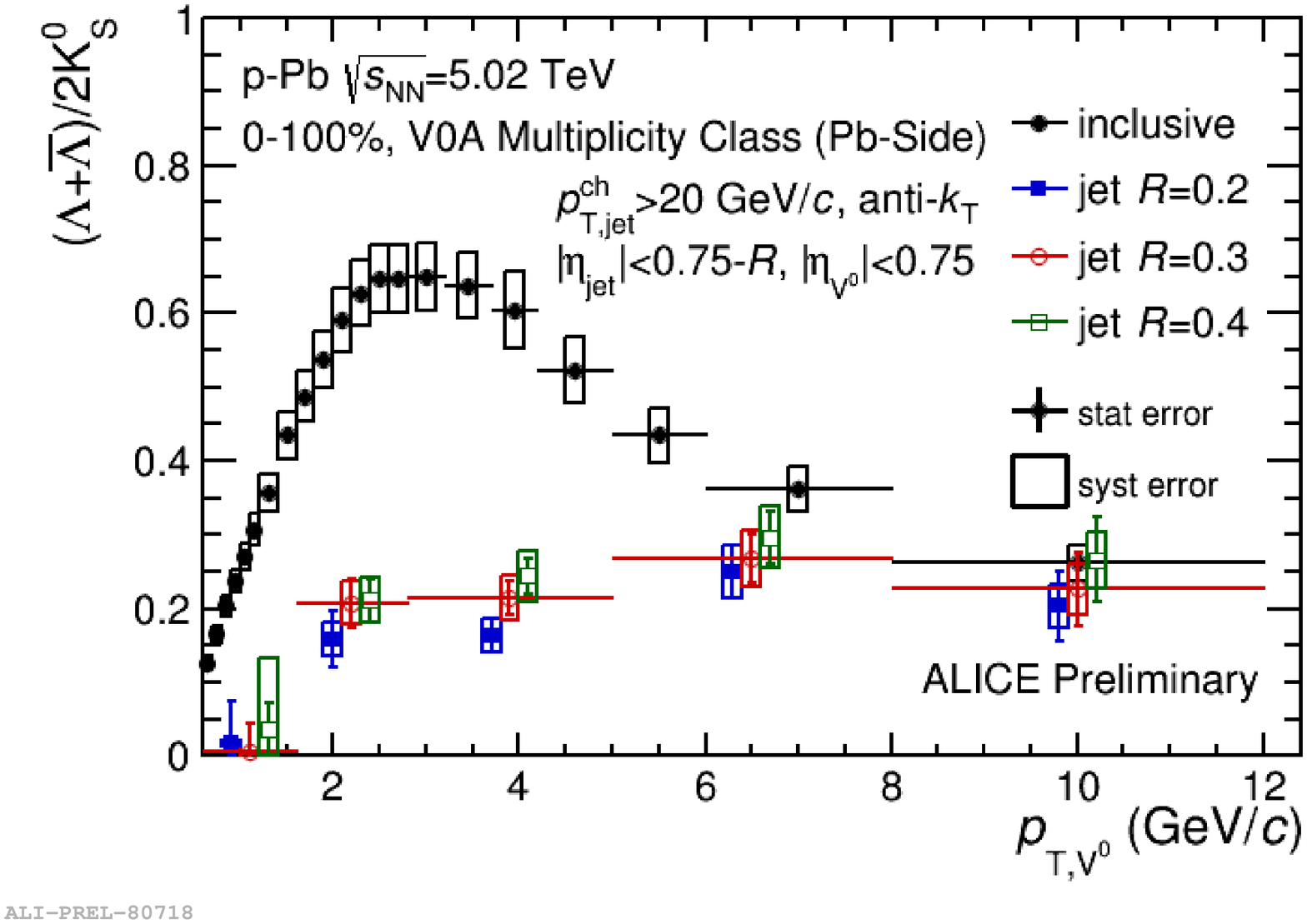}
\caption{\label{FigLKs4} Ratio of $p_{\mathrm{T}}$ spectra of $\lamb+\bar{\lamb}$ and $\particle{K}^{0}_{\mathrm{S}}$ measured in minimum bias p--Pb collisions at $\sqrt{s_{\mathrm{NN}}}=5.02$~TeV.
Black circles show the ratio obtained from  inclusive spectra.  Color markers 
show the ratio measured in charged anti-$k_{\mathrm{T}}$ jets for
 three sizes of the resolution parameter $R$.
Charged jets have transverse momentum larger than $20$~GeV/$c$. 
}
\end{minipage} 
\end{figure}

Using $\lamb$ and $\particle{K}^{0}_{\mathrm{S}}$  
particles allows to extend particle identification to higher transverse momentum
when compared to the $\particle{p}/\pion$ case.
Figure~\ref{FigLKs1} shows the ratio of inclusive spectra of  $\lamb$ and $\particle{K}^{0}_{\mathrm{S}}$ 
as measured by the ALICE experiment in Pb--Pb collisions at $\sqrt{s_{\mathrm{NN}}}=2.76$~TeV
 and in \particle{pp} at $\sqrt{s}=0.9$~TeV and 7~TeV \cite{A1}.
While \particle{pp} and  peripheral Pb--Pb collisions 
have compatible $\lamb/\particle{K}^{0}_{\mathrm{S}}$ ratios,
 in more central Pb--Pb collisions,  a gradually increasing enhancement in the range 2--7 GeV/$c$ is observed.
Above 7~GeV/$c$, all data sets tend to pp which indicates that  the 
 dominant mechanism for particle production in this region is the jet fragmentation.
On the other hand, the evolution of the ratio below 2 GeV/$c$ with $p_{\mathrm{T}}$ and centrality  is well
 reproduced by a hydro model calculation \cite{A1,s1,s2,s3}.  
The underlying mechanism of particle production in the range  2--7 GeV/$c$ 
is still a matter of discussion  \cite{AOV2,Br,Gy,We} 
but note that the signature of baryon anomaly is 
seen also in the \particle{p}--Pb at $\sqrt{s_{\mathrm{NN}}}=5.02$~TeV 
system \cite{A2}, see Fig.~\ref{FigLKs2}. 
 In these small systems, flow-like patterns might emerge due to the color reconnection as suggested by \cite{AOV1}.

A natural question to be asked is whether the baryon anomaly is produced only in the bulk 
or whether some in-medium modified jet fragmentation also contributes.
An answer to this question was searched for in the Pb--Pb and p--Pb data in parallel.
The analysis required to divide $\particle{V}^{0}$ candidates into two groups:
i) $\particle{V}^{0}$s that can be associated with a jet and ii) $\particle{V}^{0}$ that come from the underlying event.
Charged jets were reconstructed using the anti-$k_{\mathrm{T}}$ algorithm 
with resolution parameter $R=0.2$, 0.3 and 0.4.
In order to have in the Pb--Pb system a better separation between 
the combinatorial background jets that are composed of underlying event particles only 
and the real jets that contain particles produced by parton fragmentation,
 jets were required to contain a leading track 
with $p_{\mathrm{T}}$ larger than 5~GeV/$c$. Note that such condition makes the selected sample of jets more biased as will be discussed in the next section.

$\particle{V}^{0}$ reconstruction was based on charged decay channels $\lamb \rightarrow \particle{p} + \pion^{-}$
and $\particle{K}^{0}_{\mathrm{S}}\rightarrow \pion^{+} + \pion^{-}$ and employed topological cuts. 
As the decay daughters of $\particle{V}^{0}$ particles are secondary tracks they were excluded from  
the jet reconstruction procedure.
The decision to associate $\lamb$  or $\particle{K}^{0}_{\mathrm{S}}$ candidate with a given jet was done
based on their mutual angular distance,
$$\sqrt{ \Delta\varphi_{\mathrm{jet,V^{0}}}^{2} + \Delta\eta_{\mathrm{jet,V^{0}}}^{2}} < R\,.$$
Here $\Delta\varphi_{\mathrm{jet,V}^{0}}$ ($\Delta\eta_{\mathrm{jet,V^{0}}}$)
denote the azimuthal angle (pseudorapidity) distance 
between the $\particle{V}^{0}$ candidate momentum and the jet axis and 
$R$ is the jet resolution parameter. 
The extracted  yield of $\particle{V}^{0}$ candidates in jets was corrected for the expected contribution of $\particle{V}^{0}$ from the 
underlying event and for the reconstruction efficiency.
In addition, the yield of $\lamb$ particles was corrected for a contribution coming from the feed down
 of $\particle{\partgr{X}}$ cascades.

Figures~\ref{FigLKs3} and \ref{FigLKs4} compare the
$\left(\lamb+\bar{\lamb}\right)/2\particle{K}^{0}_{\mathrm{S}}$ 
ratio obtained from the inclusive  $\particle{V}^{0}$s and from  $\particle{V}^{0}$s associated with jets
with $p_{\mathrm{T,jet}}^{\mathrm{ch}}>20$~GeV/$c$ in central Pb--Pb collisions 
at $\sqrt{s_{\mathrm{NN}}}=2.76$~TeV and
 in two centrality bins of  \particle{p}--Pb collisions at $\sqrt{s_{\mathrm{NN}}}=5.02$~TeV.
In both systems, the ratio measured in jets is significantly lower than the inclusive one. 
This observation thus supports the picture that fragmentation of  hard partons 
producing charged jets with $p_{\mathrm{T,jet}}^{\mathrm{ch}}>20$~GeV/$c$ does not contribute to the 
observed baryon anomaly. A similar conclusion was reported also earlier for $\particle{p}/\pion{}$ ratio using two particle correlations \cite{Misa}.   
The source of the baryon anomaly thus seems to be intimately connected with 
 partons having smaller transverse momenta \cite{AOV2}.  

\section{h-jet correlation measurement}
The multiplicity of hadrons in a central heavy-ion collision is large. 
 In such an environment, the jet reconstruction algorithm often clusters together just
 soft bulk particles. As a result  an artificial jet is created. 
One possibility how to suppress the number of these artificial jets is to
 require that the reconstructed jet contains at least one constituent
 with $p_{\mathrm{T}}$ above some preset threshold. 
This condition, however, imposes a bias on the jet fragmentation
which can be unwanted especially in the situation
when we look for quenched jets. 

Hadron-jet coincidence measurements offer an  elegant way how to overcome this problem. 
The approach which we will discuss allows to  remove 
the contribution of combinatorial background jets including multi-parton
interaction without imposing fragmentation bias on the reconstructed jet. 
The method is data driven and can be used also for jets with large $R$ and low $p_{\mathrm{T}}$.
The detailed description can be found in \cite{ex5,ex6}.

\begin{figure}[ht]
\begin{minipage}{18pc}
\includegraphics[width=18pc]{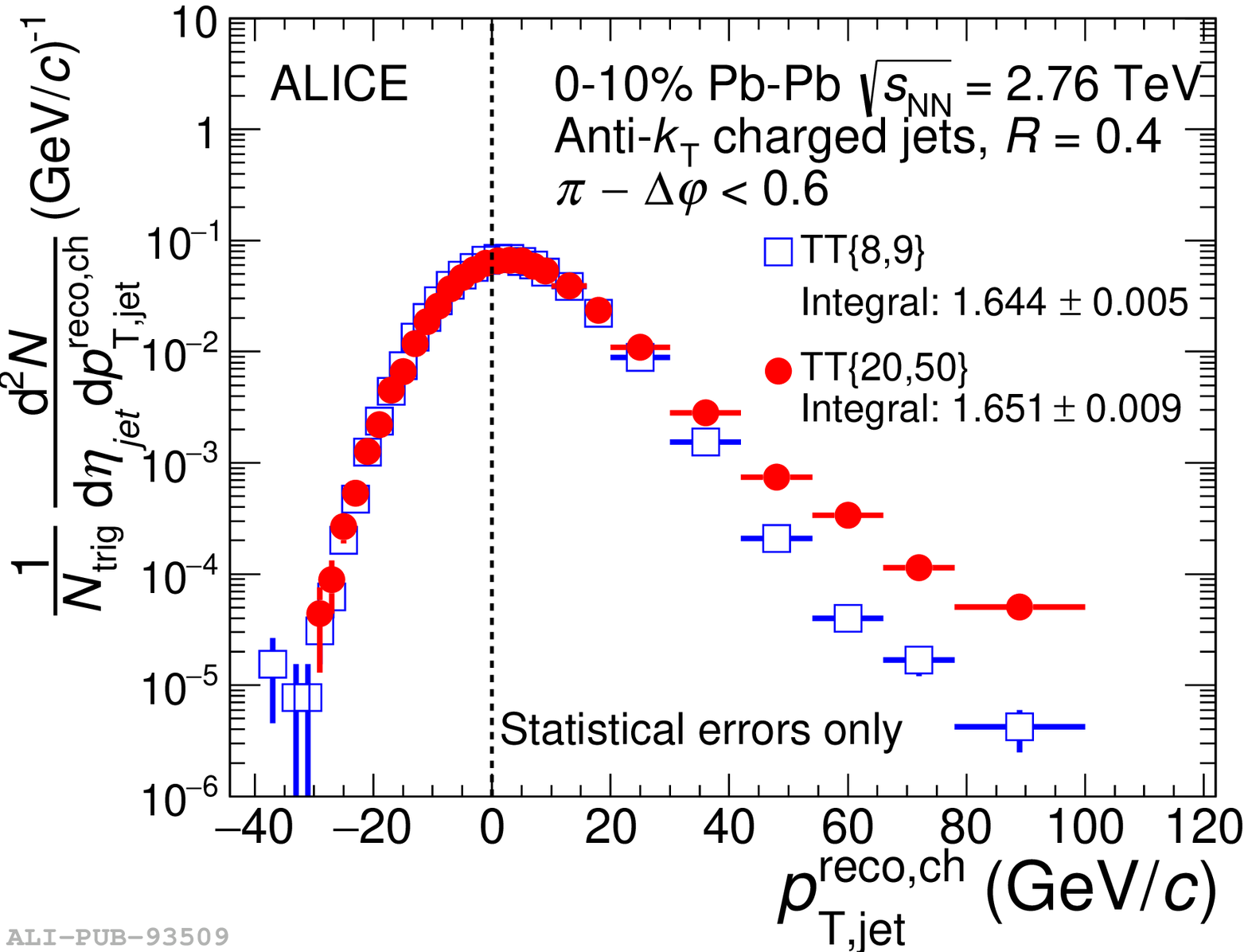}
\caption{\label{FigHJ1} Semi-inclusive distributions of jets recoiling from a hadron trigger
   with 
  $p_{\mathrm{T,trig}} \in \{20,50\}$~GeV/$c$ (red circles) and 
$p_{\mathrm{T,trig}} \in \{8,9\}$~GeV/$c$ (blue squares)  for 0--10\% 
central Pb--Pb collisions at $\sqrt{s_{\mathrm{NN}}}=2.76$~TeV.}
\end{minipage}\hspace{2pc}%
\begin{minipage}{18pc}
\includegraphics[width=18pc]{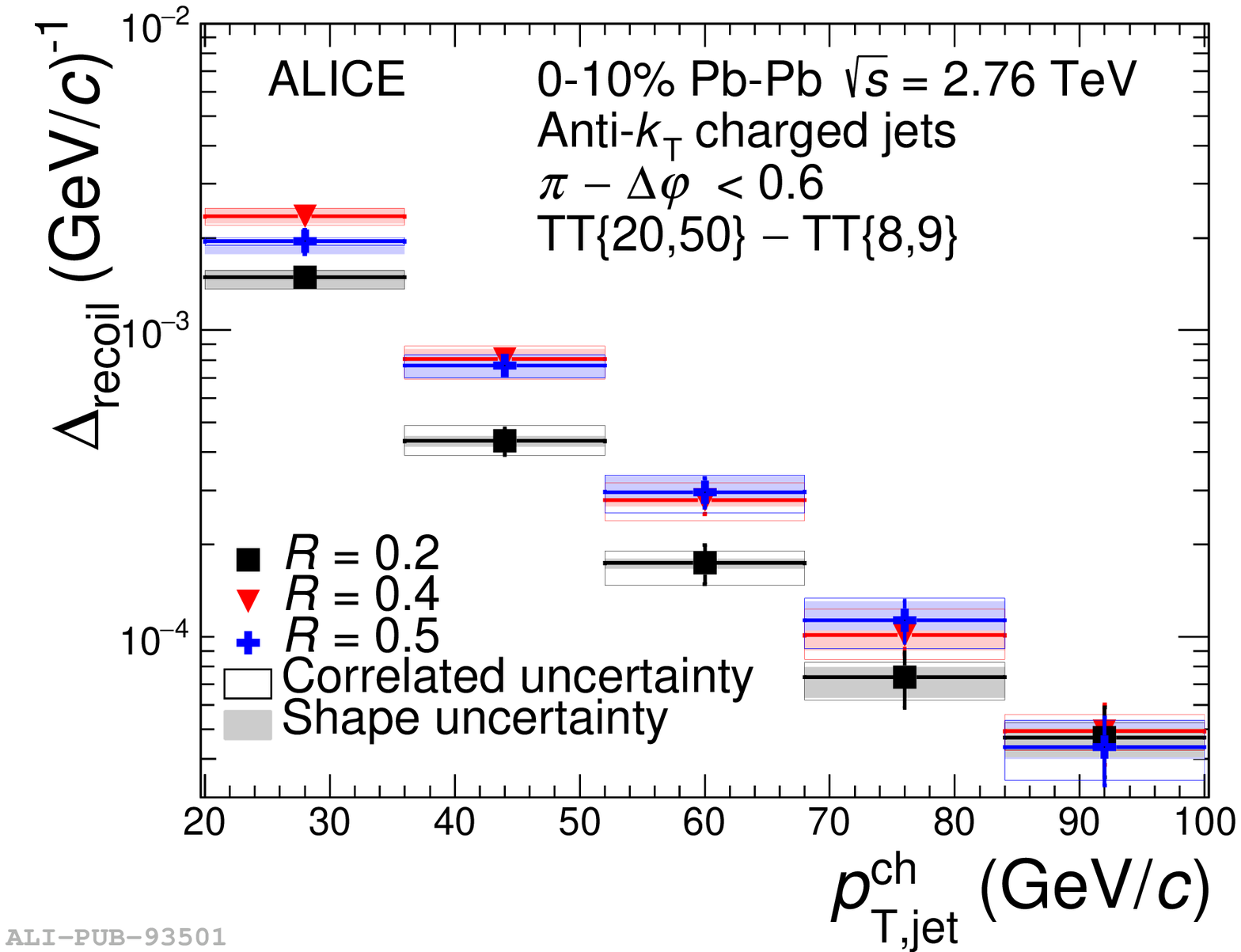}
\caption{\label{FigHJ2} Fully corrected $\Delta_{\mathrm{recoil}}$ distributions measured for 
  0--10\% central Pb--Pb collisions at $\sqrt{s_{\mathrm{NN}}}=2.76$~TeV for
  anti-$k_{\mathrm{T}}$  jets with $R=0.2$, 0.4 and 0.5. 
  The boxes indicate correlated and shape systematic uncertainties, 
   see \cite{ex6} for more details.
 }
\end{minipage} 
\end{figure}

The basic steps of the analysis can be summarized as follows. We select events that contain 
a high-$p_{\mathrm{T}}$ hadron (trigger track).
The presence of a high-$p_{\mathrm{T}}$ hadron identifies collisions where 
a hard scattering occurred.
In these events, we analyze jets that recoil nearly back to back in azimuth w.r.t. to the trigger track;
e.g., in this analysis, it was requested that the opening angle between the trigger track 
and the recoiling jet is larger than $\pi-0.6$~rad in azimuth.
Figure~\ref{FigHJ1}  shows a comparison of two per trigger  normalized 
background density corrected semi-inclusive $p_{\mathrm{T}}$ distributions of recoil jets
 associated to exclusive trigger $p_{\mathrm{T}}$ bins,
  $p_{\mathrm{T,trig}} \in \{20,50\}$~GeV/$c$ and $p_{\mathrm{T,trig}} \in \{8,9\}$~GeV/$c$.
For brevity from now on $p_{\mathrm{T,trig}}$ will be labeled as TT.
In general, a trigger track with larger $p_{\mathrm{T}}$  comes on average from 
a hard scattering process with larger $Q^2$, 
therefore also the corresponding  recoil jet $p_{\mathrm{T}}$ spectrum is harder.
Nevertheless, in the region where the mean background density corrected jet $p_{\mathrm{T}}$ 
  is around or below zero, both spectra turn out to be nearly identical.
This part of the distribution is dominated by accidental combinations
of the trigger track with uncorrelated combinatorial background jets.
The data suggests that the number of such combinations 
is largely independent of TT bin.

Assuming that the number of combinatorial background jets associated with a trigger track 
is independent of trigger track $p_{\mathrm{T}}$, it is possible to introduce the following observable
\begin{equation}
 \Delta_\mathrm{recoil} = \frac{1}{N_\mathrm{trig}} 
         \frac{{\mathrm d}^{2}N_\mathrm{jet}}{{\mathrm d}p_{\mathrm{T,jet}}^{\mathrm{ch}}{\mathrm d}\eta_{\mathrm{jet}}}\bigg|_{\mathrm{TT}\{20,50\}} - 
         \frac{1}{N_\mathrm{trig}}
        \frac{{\mathrm d}^{2}N_\mathrm{jet}}{{\mathrm d}p_{\mathrm{T,jet}}^{\mathrm{ch}}{\mathrm d}\eta_{\mathrm{jet}}}\bigg|_{\mathrm{TT}\{8,9\}}
\end{equation}
where $N_\mathrm{trig}$ is the number of trigger tracks in a given TT bin.
In this observable the contribution of combinatorial background jets is removed.  
Further, let us point out that $\Delta_{\mathrm{recoil}}$ has a direct link to theory,
since the  per trigger yield of recoil jets can be expressed in terms of the cross-section to
produce a high-$p_\mathrm{T}$ hadron and the cross section to produce 
a high-$p_\mathrm{T}$ hadron together with a jet, i.e., 
\begin{equation}
{1\over N_{\mathrm{trig}}^{\mathrm{AA}}} { \mathrm{d}^{2} N_{\mathrm{jet}}^{\mathrm{AA}} \over \mathrm{d} p_{\mathrm{T,jet}}^{\mathrm{ch}} \mathrm{d} \eta_{\mathrm{jet}}} \bigg|_{p_{\mathrm{T,trig}} \in \mathrm{TT}} = \left( {1 \over \sigma^{\mathrm{AA}\rightarrow \mathrm{h+X}}}  \cdot
{ \mathrm{d}^{2} \sigma^{\mathrm{AA} \rightarrow \mathrm{h+jet+X}}  \over \mathrm{d}p_{\mathrm{T,jet}}^{\mathrm{ch}} \mathrm{d}\eta_{\mathrm{jet}}}\right)\bigg|_{p_{\mathrm{T,h}}\in \mathrm{TT}}
\end{equation}

The requirement to have a high-$p_{\mathrm{T}}$ trigger in event introduces 
also some interesting biases on the selected sample of events and recoil jets.
E.g., the parton that produces the trigger particle is more biased to be close to the surface
of the collision zone \cite{Renk}, therefore the recoiling jet will have on average longer
 path length through the medium.  Events containing a hard particle are also more biased
to have larger number of participants or in other words to be more central. 
In both TT bins that are used this bias is the same. 

\begin{figure}[ht]
\includegraphics[width=16pc]{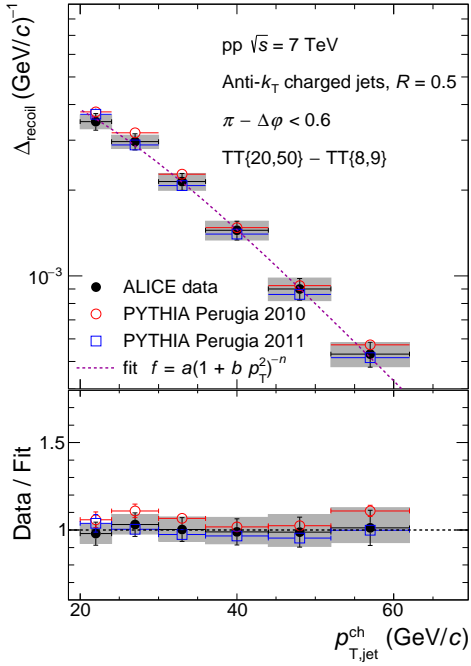}\hspace{2pc}%
\begin{minipage}[b]{12pc}\caption{\label{HJ3} Upper panel: 
$\Delta_{\mathrm{recoil}}$ distributions 
corresponding to anti-$k_\mathrm{T}$ jets with $R=0.5$ in \particle{pp} collisions at $\sqrt{s}=7$~TeV. 
Comparison of measured data and PYTHIA Perugia 2010 and 2011 calculation. 
Systematic uncertainties on the measured data points are 
shown as gray boxes. The measured data are fit with a smooth function.
Bottom panel: ratios of the data sets to the fit. }  
\end{minipage}
\end{figure}

In Figure~\ref{FigHJ2}, we present $\Delta_{\mathrm{recoil}}$ 
distributions for several values of 
resolution parameter. The distributions are 
corrected for background fluctuations and detector effects. 
The medium-induced modification of jet fragmentation can be then
searched for by means of the ratio
\begin{equation} 
    \Delta I_{\mathrm{AA}} = \frac{ \Delta_{\mathrm{recoil}}^{\mathrm{PbPb}}}{\Delta_{\mathrm{recoil}}^{\mathrm{pp}}}
\end{equation} 
where 
the $\Delta_{\mathrm{recoil}}$ distribution measured in Pb--Pb is divided by 
the reference $\Delta_{\mathrm{recoil}}$ distribution from \particle{pp} collisions 
at the same center of mass energy per nucleon-nucleon pair.
As ALICE data from \particle{pp} at $\sqrt{s}=2.76$~TeV have poor statistics we 
use  a reference spectrum generated by the PYTHIA Perugia 2010 tune \cite{PY}.
To test the reliability of the PYTHIA prediction, 
  we have cross-checked the PYTHIA calculation with the measured  $\Delta_{\mathrm{recoil}}$ 
 spectrum obtained from \particle{pp} at $\sqrt{s}=7$~TeV data,
see Fig.~\ref{HJ3}. The \particle{pp} analysis closely followed what was done for Pb--Pb.
 In general we can say that PYTHIA Perugia tunes provide 
predictions that are compatible with the measured data within 
the quoted statistical and systematic uncertainties.

\begin{figure}[ht]
\begin{minipage}{18pc}
\includegraphics[width=18pc]{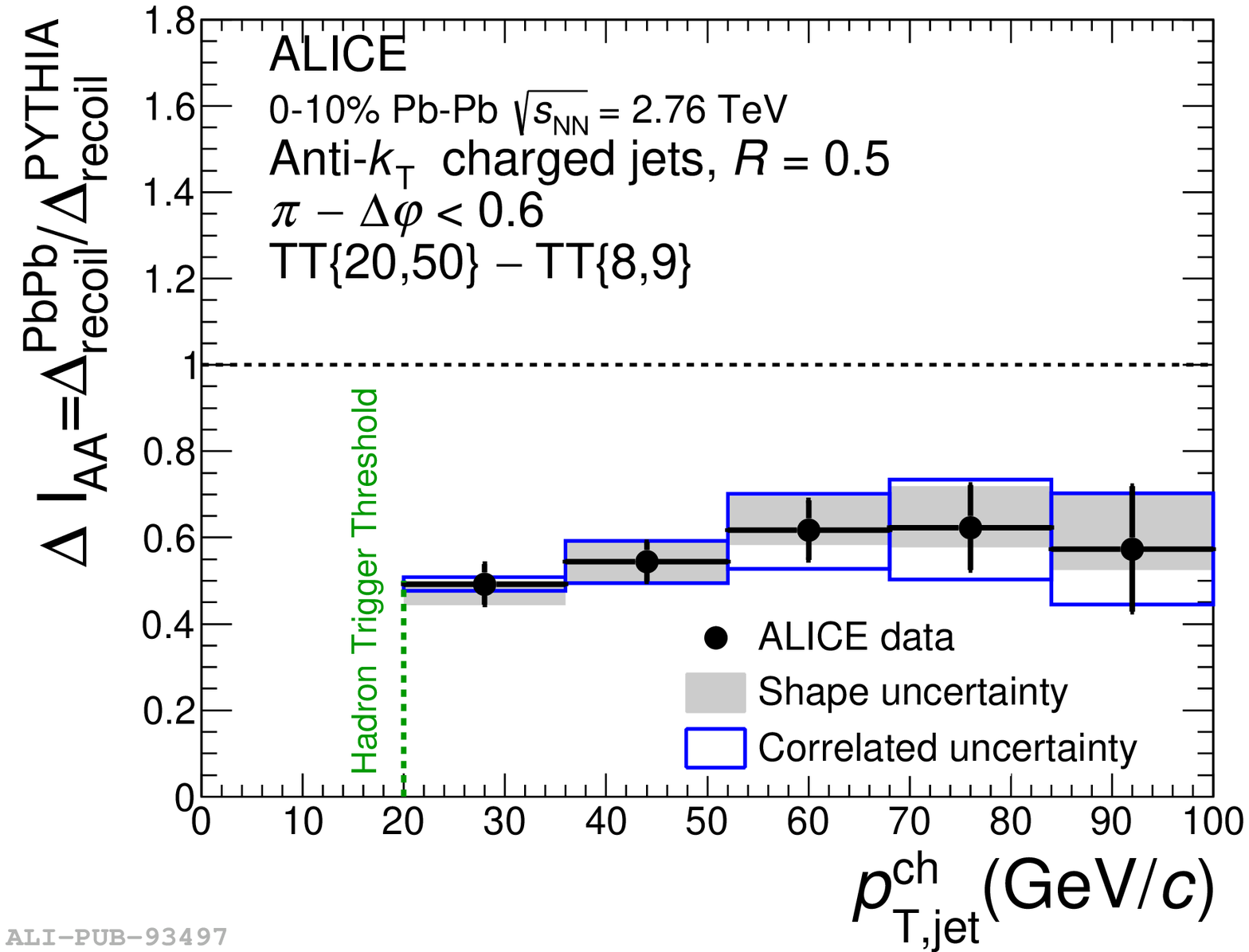}
\caption{\label{FigHJ4}   $\Delta I_{\mathrm{AA}}$, the ratio of $\Delta_{\mathrm{recoil}}$ 
measured in central Pb--Pb and pp collisions at $\sqrt{s_{\mathrm{NN}}}=2.76$~TeV. 
 $\Delta_{\mathrm{recoil}}$ for pp collision was calculated with PYTHIA Perugia 10. 
 The data corresponds to charged anti-$k_{\mathrm{T}}$ jets with $R=0.5$.
 }
\end{minipage}\hspace{2pc}%
\begin{minipage}{18pc}
\includegraphics[width=18pc]{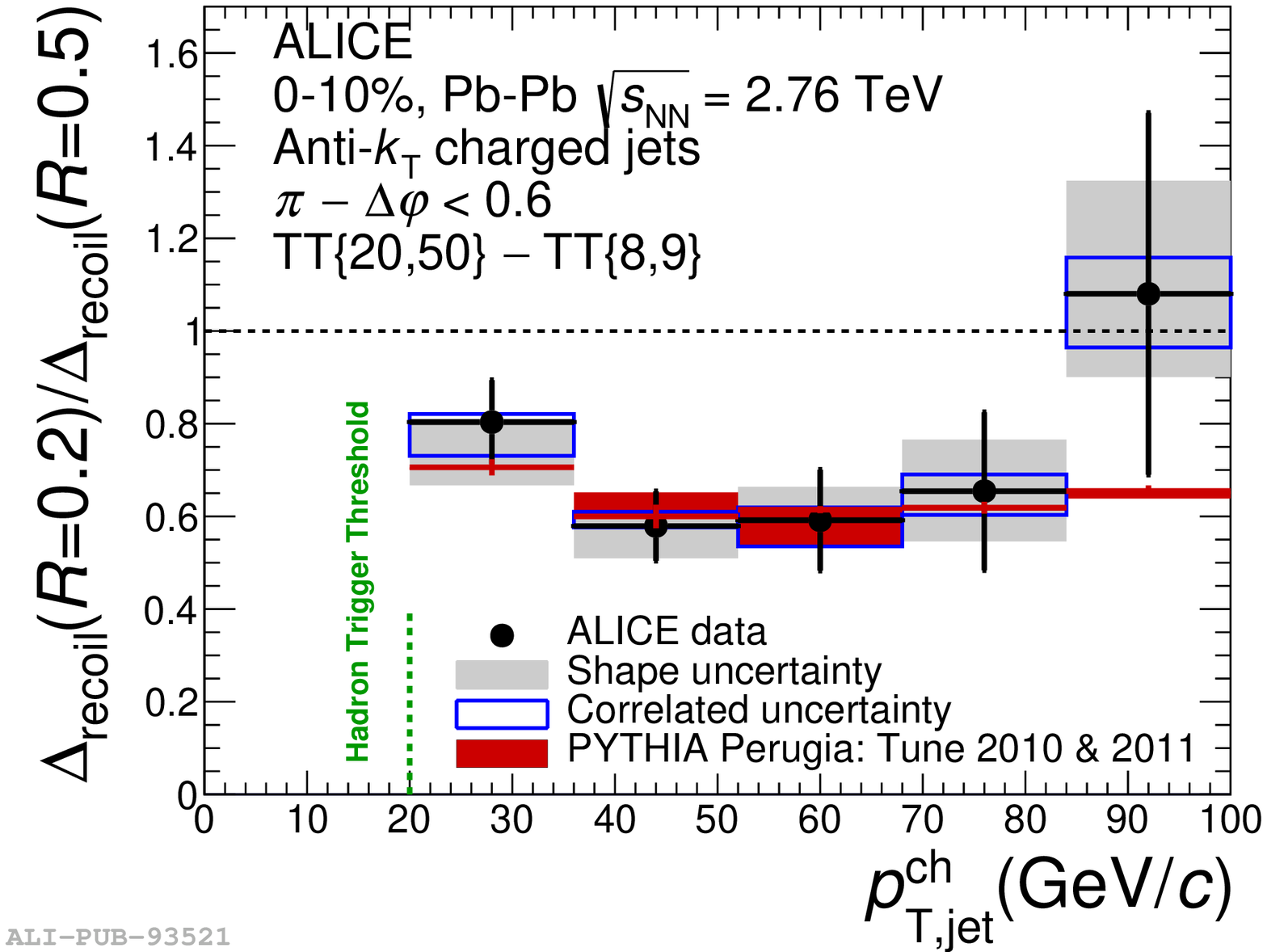}
\caption{\label{FigHJ5} 
\label{FigHJ6}  Ratio of $\Delta_{\mathrm{recoil}}$ distributions
corresponding to  charged anti-$k_{\mathrm{T}}$ jets with $R=0.5$ and $R=0.2$. 
The black circles show the data measured in central Pb--Pb at $\sqrt{s_{\mathrm{NN}}}=2.76$~TeV.
An estimate of the ratio in \particle{pp} 
was obtained using PYTHIA Perugia tunes and is represented by the red bend.
The width of the red bend  is given by the difference between the Perugia 2010 and 2011 tune. 
}
\end{minipage} 
\end{figure}

Figure~\ref{FigHJ4} shows the $\Delta I_{AA}$ obtained with PYTHIA Perugia 10 
reference for anti-$k_{\mathrm{T}}$ jets with $R=0.5$.
The ratio exhibits a deficit in the yield of recoil jets.
The magnitude of this suppression for jets with $R=0.2$ and $0.4$ is similar.

The $\Delta_{\mathrm{recoil}}$ observable allows also to look for possible  medium-induced
modification of the jet shape. Figure~\ref{FigHJ5}  presents
 the ratio of $\Delta_{\mathrm{recoil}}$ spectra of jets obtained for different $R$.
The measured data are compared with vacuum \particle{pp} trend  predicted by PYTHIA.
Within the quoted errors
there is no evidence for significant energy redistribution w.r.t. PYTHIA pp data.
This also means that there is no evidence for intra-jet broadening up to $R=0.5$.

\section{Summary}
 Jets as well as other hard probes  bring us information
 about the early states of matter produced in a collision of two nuclei.
 The picture of jet-medium interaction would, however, be
  incomplete without the precise understanding of jet production and properties 
  in elementary reactions such as \particle{pp}.
  The measurement of particle type dependent jet fragmentation 
  by ALICE in \particle{pp} collisions at $\sqrt{s}=7$~TeV
thus helps to constraint available fragmentation models and event generators 
  that are on the market.

 Complex final state interaction gives rise to the baryon anomaly
which is observed both in Pb--Pb and p--Pb systems at LHC energies. 
  Measurements of $\lamb/\particle{K}^{0}_{\mathrm{S}}$  ratio
  done with inclusive particles and jet fragments 
  indicate that fragmentation of partons producing charged jets with 
  $p_{\mathrm{T,jet}}^{\mathrm{ch}}>20$~GeV/$c$ 
 is not a relevant source for the observed anomaly
 which thus seems to emerge from processes where the transferred $Q^2$  
 was lower. 

Hadron-jet correlation measurements offer a new variety of observables well suited  to study 
also low-$p_{\mathrm{T}}$ jets 
with large resolution parameter without inducing a fragmentation bias.
PYTHIA Perugia tunes seem to provide reliable prediction for the measured 
  h-jet observables in \particle{pp} collisions, such as $\Delta_{\mathrm{recoil}}$. 
  In Pb--Pb system, we see a suppression of the recoil jet yield 
  but without a sign of intra-jet broadening for charged anti-$k_\mathrm{T}$ jets
  up to resolution parameter $R=0.5$.

\section*{ Acknowledgments}
The work has been supported by the MEYS grant CZ.1.07/2.3.00/20.0207 of the European
Social Fund (ESF) in the Czech Republic: Education for Competitiveness Operational
Programme (ECOP) and by the grant LG 13031 of the Ministry of Education of the Czech
Republic.

\end{document}